\documentclass{article}%
\usepackage{amsmath}%
\usepackage{amsfonts}%
\usepackage{amssymb}%
\usepackage{graphicx}
\newtheorem{theorem}{Theorem}
\newtheorem{acknowledgement}[theorem]{Acknowledgement}

\newtheorem{remark}[theorem]{Remark}

\begin{document}

\title{Quaternionic equation for electromagnetic fields in inhomogeneous media}
\author{Vladislav V. Kravchenko\\Depto. de Telecomunicaciones,\\Escuela Superior de Ingenier\'{\i}a Mec\'{a}nica y El\'{e}ctrica,\\Instituto Polit\'{e}cnico Nacional,\\C.P.07738, D.F., MEXICO\\e-mail: vkravche@maya.esimez.ipn.mx}
\date{October 25, 2001}
\maketitle

\begin{abstract}
We show that the Maxwell equations for arbitrary inhomogeneous media are
equivalent to a single quaternionic equation which can be considered as a
generalization of the Vekua equation for generalized analytic functions.

\end{abstract}

\section{Introduction}

Quaternionic reformulation of the Maxwell equations in a vacuum is quite well
known (see, for example, \cite{BKT}, \cite{GurseyTze}, \cite{Imaeda},
\cite{KSbook}). The system of Maxwell equations%
\begin{align}
\operatorname{div}\mathbf{D}  & =\operatorname{div}\mathbf{B}=0\nonumber\\
& \label{1}\\
\operatorname{rot}\mathbf{E}  & =-\partial_{t}\mathbf{B},\qquad
\operatorname{rot}\mathbf{H}=\partial_{t}\mathbf{D},\nonumber
\end{align}
where $D=\varepsilon_{0}E$ and $B=\mu_{0}H$ ($\varepsilon_{0}$ and $\mu_{0}$
are the permittivity and permeability of free space), is equivalent to the
quaternionic equation%
\begin{equation}
(\frac{1}{c_{0}}\partial_{t}+iD)\overrightarrow{f}=0,\label{2}%
\end{equation}
where $c_{0}$ is the speed of light in a vacuum, $D$ is the Moisil-Theodoresco
operator (see the definition below) and $\overrightarrow{f}$ is a purely
vectorial complex quaternion. The equivalence between (\ref{1}) and (\ref{2})
can be established by putting%
\[
\operatorname{Re}\overrightarrow{f}=\sqrt{\varepsilon_{0}}E\qquad
\text{and\qquad}\operatorname{Im}\overrightarrow{f}=\sqrt{\mu_{0}}H.
\]
The quaternionic approach to Maxwell's equations for homogeneous media was
intensively used in a number of works (e.g., \cite{KKO}, \cite{KKR},
\cite{Krdep}, \cite{KSbook}), but even the question as to how to write the
Maxwell equations for arbitrary inhomogeneous media in a compact quaternionic
form remained open until recently (an attempt in this direction can be found
in \cite[Section 4.5]{GS2}). In \cite{KrZAAnew} such a reformulation was
proposed in the case of a time-harmonic electromagnetic field and in
\cite{KrGraz} for the time-dependent case. Here we make one additional step
which leads us to Maxwell's system in the form of a single quaternionic
equation, which can be considered as a generalization of the well known in
complex analysis Vekua equation describing generalized analytic functions
\cite{Vekua}.

\section{Preliminaries}

We will consider continuously differentiable functions of four variables
$(t,x_{1},x_{2},x_{3})$ with values in the algebra of complex quaternions
$\mathbb{H}(\mathbb{C})$. By $D$ we denote the operator $D=\sum_{k=1}^{3}%
i_{k}\partial_{k}$. Here $\partial_{k}=\frac{\partial}{\partial x_{k}}$ and
$i_{k}$ are the imaginary quaternionic units. Let us notice the following
property of the operator $D$. Let $\varphi$ be a scalar complex function and
$g$ be an $\mathbb{H}(\mathbb{C})$-valued function. Then
\begin{equation}
D[\varphi\cdot g]=D[\varphi]\cdot g+\varphi\cdot D[g].\label{3}%
\end{equation}
Taking into account that $D[\varphi]=\operatorname{grad}\varphi=i_{1}%
\partial_{1}\varphi+i_{2}\partial_{2}\varphi+i_{3}\partial_{3}\varphi$ and
assuming that $\varphi$ is different from zero we can rewrite (\ref{3}) in the
form%
\begin{equation}
(D+\frac{\operatorname{grad}\varphi}{\varphi})g=\frac{1}{\varphi}%
D[\varphi\cdot g].\label{4}%
\end{equation}
We will use the following notations for the operators of multiplication from
the left-hand side and from the right-hand side%
\[
^{\alpha}Mg:=\alpha\cdot g\qquad\text{and\qquad}M^{\alpha}g:=g\cdot\alpha,
\]
where $\alpha\in\mathbb{H}(\mathbb{C})$. The usual complex conjugation we
denote by ``*''. Vectors from $\mathbb{C}^{3}$ are identified with purely
vectorial complex quaternions. Note that for an $\mathbb{H}(\mathbb{C}%
)$-valued function $g=g_{0}+\overrightarrow{g}$ the action of the operator $D$
can be represented as follows%
\[
Dg=-\operatorname*{div}\overrightarrow{g}+\operatorname*{grad}g_{0}%
+\operatorname*{rot}\overrightarrow{g}.
\]

\section{Maxwell equations}

We assume that the relative permittivity $\varepsilon_{r}$ and the relative
permeability $\mu_{r}$ of the material are differentiable functions of
coordinates $\varepsilon_{r}=\varepsilon_{r}(x_{1},x_{2},x_{3})$ and $\mu
_{r}=\mu_{r}(x_{1},x_{2},x_{3})$. The permittivity and the permeability of the
medium are introduced as follows%
\[
\varepsilon=\varepsilon_{0}\varepsilon_{r}\text{\qquad and\qquad}\mu=\mu
_{0}\mu_{r}.
\]
Then Maxwell's equations for an inhomogeneous medium have the form%
\begin{equation}
\operatorname{rot}\mathbf{H}=\varepsilon\partial_{t}\mathbf{E}+\mathbf{j,}%
\label{Min1}%
\end{equation}%
\begin{equation}
\operatorname{rot}\mathbf{E}=-\mu\partial_{t}\mathbf{H},\label{Min2}%
\end{equation}%
\begin{equation}
\operatorname{div}(\varepsilon\mathbf{E)}=\mathbf{\rho},\label{Min3}%
\end{equation}%
\begin{equation}
\operatorname{div}\mathbf{(}\mu\mathbf{H)}=0,\label{Min4}%
\end{equation}
where all the magnitudes are real. Equations (\ref{Min3}) and (\ref{Min4}) can
be written as follows%
\[
\operatorname*{div}\mathbf{E}+<\frac{\operatorname{grad}\varepsilon
}{\varepsilon},\mathbf{E}>=\frac{\mathbf{\rho}}{\varepsilon}
\]
and%
\[
\operatorname*{div}\mathbf{H}+<\frac{\operatorname{grad}\mu}{\mu}%
,\mathbf{H}>=0,
\]
where $<\cdot,\mathbf{\cdot}>$ denotes the usual scalar product. Combining
these equations with (\ref{Min1}) and (\ref{Min2}) we obtain the Maxwell
system in the form%
\begin{equation}
D\mathbf{E}=<\frac{\operatorname{grad}\varepsilon}{\varepsilon},\mathbf{E}%
>-\mu\partial_{t}\mathbf{H}-\frac{\mathbf{\rho}}{\varepsilon}\label{Min11}%
\end{equation}
and%
\begin{equation}
D\mathbf{H}=<\frac{\operatorname{grad}\mu}{\mu},\mathbf{H}>+\varepsilon
\partial_{t}\mathbf{E}+\mathbf{j}.\label{Min12}%
\end{equation}
Let us make a simple observation: the scalar product of two vectors
$\overrightarrow{p}$ and $\overrightarrow{q}$ can be represented as follows%
\[
<\overrightarrow{p},\overrightarrow{q}>=-\frac{1}{2}(^{\overrightarrow{p}%
}M+M^{\overrightarrow{p}})\overrightarrow{q}.
\]
Using this fact, from (\ref{Min11}) and (\ref{Min12}) we obtain the pair of
equations%
\begin{equation}
(D+\frac{1}{2}\frac{\operatorname{grad}\varepsilon}{\varepsilon}%
)\mathbf{E}=-\frac{1}{2}M^{\frac{\operatorname{grad}\varepsilon}{\varepsilon}%
}\mathbf{E}-\mu\partial_{t}\mathbf{H}-\frac{\mathbf{\rho}}{\varepsilon
}\label{Min21}%
\end{equation}
and%
\begin{equation}
(D+\frac{1}{2}\frac{\operatorname{grad}\mu}{\mu})\mathbf{H}=-\frac{1}%
{2}M^{\frac{\operatorname{grad}\mu}{\mu}}\mathbf{H}+\varepsilon\partial
_{t}\mathbf{E}+\mathbf{j}.\label{Min22}%
\end{equation}
Note that
\[
\frac{1}{2}\frac{\operatorname{grad}\varepsilon}{\varepsilon}%
=\frac{\operatorname{grad}\sqrt{\varepsilon}}{\sqrt{\varepsilon}}.
\]
Then using (\ref{4}), equation (\ref{Min21}) can be rewritten in the following
form%
\begin{equation}
\frac{1}{\sqrt{\varepsilon}}D(\sqrt{\varepsilon}\cdot\mathbf{E)}%
+\mathbf{E}\cdot\overrightarrow{\varepsilon}=-\mu\partial_{t}\mathbf{H}%
-\frac{\mathbf{\rho}}{\varepsilon},\label{Min31}%
\end{equation}
where%
\[
\overrightarrow{\varepsilon}:=\frac{\operatorname{grad}\sqrt{\varepsilon}%
}{\sqrt{\varepsilon}}.
\]
Analogously, (\ref{Min22}) takes the form%
\begin{equation}
\frac{1}{\sqrt{\mu}}D(\sqrt{\mu}\cdot\mathbf{H)}+\mathbf{H}\cdot
\overrightarrow{\mu}=\varepsilon\partial_{t}\mathbf{E}+\mathbf{j,}%
\label{Min32}%
\end{equation}
where%
\[
\overrightarrow{\mu}:=\frac{\operatorname{grad}\sqrt{\mu}}{\sqrt{\mu}}.
\]
Introducing the notations%
\[
\overrightarrow{\mathcal{E}}:=\sqrt{\varepsilon}\mathbf{E,\qquad
}\overrightarrow{\mathcal{H}}:=\sqrt{\mu}\mathbf{H}
\]
and multiplying (\ref{Min31}) by $\sqrt{\varepsilon}$ and (\ref{Min32}) by
$\sqrt{\mu}$ we arrive at the equations%
\begin{equation}
(D+M^{\overrightarrow{\varepsilon}})\overrightarrow{\mathcal{E}}=-\frac{1}%
{c}\partial_{t}\overrightarrow{\mathcal{H}}-\frac{\mathbf{\rho}}%
{\sqrt{\varepsilon}},\label{Minq1}%
\end{equation}
and%
\begin{equation}
(D+M^{\overrightarrow{\mu}})\overrightarrow{\mathcal{H}}=\frac{1}{c}%
\partial_{t}\overrightarrow{\mathcal{E}}+\sqrt{\mu}\mathbf{j},\label{Minq2}%
\end{equation}
where $c=1/\sqrt{\varepsilon\mu}$ is the speed of propagation of
electromagnetic waves in the medium.

Equations (\ref{Minq1}) and (\ref{Minq2}) \ can be rewritten even in a more
elegant form. Consider the function%
\[
\overrightarrow{f}:=\overrightarrow{\mathcal{E}}+i\overrightarrow
{\mathcal{H}}
\]
Let us apply to it the quaternionic Maxwell operator
\[
\frac{1}{c}\partial_{t}+iD.
\]
We obtain%
\[
(\frac{1}{c}\partial_{t}+iD)\overrightarrow{f}=\frac{1}{c}\partial
_{t}\overrightarrow{\mathcal{E}}-D\overrightarrow{\mathcal{H}}+i(\frac{1}%
{c}\partial_{t}\overrightarrow{\mathcal{H}}+D\overrightarrow{\mathcal{E}}).
\]
For the real part of this expression we use equation (\ref{Minq2}) and for the
imaginary part equation (\ref{Minq1}). Then we have

\bigskip%
\begin{equation}
(\frac{1}{c}\partial_{t}+iD)\overrightarrow{f}=-i(M^{\overrightarrow
{\varepsilon}}\overrightarrow{\mathcal{E}}+iM^{\overrightarrow{\mu}%
}\overrightarrow{\mathcal{H}})-\sqrt{\mu}\mathbf{j}-\frac{i\mathbf{\rho}%
}{\sqrt{\varepsilon}}.\label{vsp4111}%
\end{equation}
Note that
\[
\overrightarrow{\mathcal{E}}=\frac{1}{2}(\overrightarrow{f}+\overrightarrow
{f}^{\ast})\qquad\text{and\qquad}\overrightarrow{\mathcal{H}}=\frac{1}%
{2i}(\overrightarrow{f}-\overrightarrow{f}^{\ast}).
\]
Hence%
\[
M^{\overrightarrow{\varepsilon}}\overrightarrow{\mathcal{E}}%
+iM^{\overrightarrow{\mu}}\overrightarrow{\mathcal{H}}=\frac{1}{2}%
(M^{(\overrightarrow{\varepsilon}+\overrightarrow{\mu})}\overrightarrow
{f}\mathbf{+}M^{(\overrightarrow{\varepsilon}-\overrightarrow{\mu}%
)}\overrightarrow{f}^{\ast}).
\]
Let us notice that
\[
\overrightarrow{\varepsilon}+\overrightarrow{\mu}=-\frac{\operatorname{grad}%
c}{c}\qquad\text{and\qquad}\overrightarrow{\varepsilon}-\overrightarrow{\mu
}=-\frac{\operatorname{grad}W}{W},
\]
where $W=\sqrt{\mu/\varepsilon}$ is the intrinsic wave impedance of the
medium. Denote%
\[
\overrightarrow{c}:=\frac{\operatorname{grad}\sqrt{c}}{\sqrt{c}}%
\qquad\text{and\qquad}\overrightarrow{W}:=\frac{\operatorname{grad}\sqrt{W}%
}{\sqrt{W}}.
\]
Then%
\[
M^{\overrightarrow{\varepsilon}}\overrightarrow{\mathcal{E}}%
+iM^{\overrightarrow{\mu}}\overrightarrow{\mathcal{H}}=-(M^{\overrightarrow
{c}}\overrightarrow{f}\mathbf{+}M^{\overrightarrow{W}}\overrightarrow{f}%
^{\ast}).
\]
From (\ref{vsp4111}) we obtain the Maxwell equations for an inhomogeneous
medium in the following form%
\begin{equation}
(\frac{1}{c}\partial_{t}+iD)\overrightarrow{f}\mathbf{-}M^{i\overrightarrow
{c}}\overrightarrow{f}\mathbf{-}M^{i\overrightarrow{W}}\overrightarrow
{f}^{\ast}=-(\sqrt{\mu}\mathbf{j}+\frac{i\mathbf{\rho}}{\sqrt{\varepsilon}%
})\label{Maxmain}%
\end{equation}
(compare with (\ref{2})). This equation is completely equivalent to the
Maxwell system (\ref{Min1})-(\ref{Min4}) and represents Maxwell's equation for
inhomogeneous media in a quaternionic form.

\begin{remark}
Equation (\ref{Maxmain}) can be considered as a generalization of the well
known in complex analysis Vekua equation describing generalized analytic
functions \cite{Vekua}. Recently in \cite{Mal98} using the L. Bers approach
\cite{Bers1}, \cite{Bers2} another quaternionic generalization of the Vekua
equation was considered. Probably some of the interesting results discussed in
\cite{Mal98} can be obtained for (\ref{Maxmain}) also. Then their physical
meaning would be of a great interest.
\end{remark}

\begin{acknowledgement}
This work was supported by CONACYT Project 32424-E, Mexico.
\end{acknowledgement}

\end{document}